\documentclass{article}
\usepackage{graphicx}  % standard LaTeX graphics tool;
\usepackage{amsmath}   % For getting proper math eqns
\usepackage[compress]{cite}
\usepackage{amssymb}   % Used for getting various symbols eg. gtrsim, lesssim etc.
\usepackage{bm} % For bold math (esp of lower greek letters)
\usepackage{dcolumn}% Align table columns on decimal point
\usepackage{color}
\usepackage{mathrsfs}
\usepackage{amsfonts}
\usepackage{varioref}
\usepackage{braket}
\usepackage{subcaption}
\usepackage{comment}

\RequirePackage[colorlinks,citecolor=blue,urlcolor=magenta,linkcolor=blue]{hyperref}
\labelformat{section}{Section #1} 
\labelformat{subsection}{Section #1} 
\labelformat{subsubsection}{Section #1}
\labelformat{subsubsubsection}{Section #1}
\labelformat{equation}{Eq.~(#1)} 
\labelformat{figure}{Fig.~#1} 
\labelformat{subfigure}{Fig.~\thefigure#1} 
\labelformat{table}{Tab.~#1} 
\labelformat{appendix}{Appendix #1}
\title{ Wave function of the Universe as a sum over eventually inflating universes}
\author{Karthik Rajeev\footnote{karthikrajeev.kr@gmail.com}$~^{1,2}$
	\\
	\\
	{$~^{1}$\small{School of Physical Sciences,}}\\
	{\small{Indian Association for the Cultivation of Science, Kolkata-700032, India}}\\
	{$~^{2}$\small{Department of Physics,}}\\
{\small{Indian Institute of Technology Bombay, Mumbai, 400076, India}}
}
\date{ }  %% This command  will supress printing the date. 
\begin{document}
	\maketitle
	%%%%%%%%%%%%%%%%%%%%%%%%%%%%%%%%%%%%%%%%%%%%%%%%%%%%%%%%%%%%%%%%%%%%%%%%%%%%%%%%%%%%%%%%%%%%%%%%%%%
	%%%%%%%%%%%%%%%%%%%%%%%%%%%%%%%%%%%%%%%%%%%%%%%%%%%%%%%%%%%%%%%%%%%%%%%%%%%%%%%%%%%%%%%%%%%%%%%%%%%
	%%%%%%%%%%%%%%%%%%%%%%%%%%%%%%%%%%%%%%%%%%%%%%%%%%%%%%%%%%%%%%%%%%%%%%%%%%%%%%%%%%%%%%%%%%%%%%%%%%%
	\begin{abstract}
		We consider a proposal to define the wave function of the Universe as a sum over spacetimes that eventually inflate. In the minisuperspace model, we explicitly show that a simple family of initial conditions, parametrized by a positive real number $a_0$, can be imposed to realise this prescription. The resulting wave function is found to be proportional to the Hartle-Hawking wave function and its dependence on $a_0$ is only through an overall phase factor. Motivated by this observation, we ask whether it is possible to analytically extend $a_0$ to an extended region $\bar{\mathcal{D}}$ in complex $a_0-$plane, while retaining the Hartle-Hawking form of the wave function. We use the condition for convergence of path integral and a recent theorem due to Kontsevich and Segal, further extended by Witten, to explicitly find $\bar{\mathcal{D}}$. Interestingly, a special point on the boundary of $\bar{\mathcal{D}}$ recovers the exact no-boundary wave function. Following that, we show that our prescription leads to a family of quantum states for the perturbations, which give rise to scale-invariant power spectra. If we demand, as an extra ingredient to our prescription, a matching condition at the ``no-boundary point'' in $\bar{\mathcal{D}}$, we converge on a unique quantum state for the perturbations. 
	\end{abstract}
	%%%%%%%%%%%%%%%%%%%%%%%%%%%%%%%%%%%%%%%%%%%%%%%%%%%%%%%%%%%%%%%%%%%%%%%%%%%%%%%%%%%%%%%%%%%%%%%%%%%
	%%%%%%%%%%%%%%%%%%%%%%%%%%%%%%%%%%%%%%%%%%%%%%%%%%%%%%%%%%%%%%%%%%%%%%%%%%%%%%%%%%%%%%%%%%%%%%%%%%%
	\section{Introduction}	
	Although the idea of describing the whole universe as a quantum system might sound like the most ambitious goal in theoretical physics, there has been tremendous progress in this direction. However, a complete understanding of this problem requires tools to handle gravity in quantum settings. Given that one does not have a realistic and consistent quantum theory of gravity, explorations on the quantum aspects of universe calls for creative and bold proposals. Two pioneering approaches in this direction, that are still relevant and remain topics of active investigations, are often referred to in the literature as the``no-boundary proposal''\cite{Hartle:1983ai} and the ``tunnelling proposal''\cite{Vilenkin:1982de}. While the specifics of these two ideas differ in important ways, the key philosophical motivation is the same; to concretely describe the creation of universe from ``nothing''.      
	
	To make progress towards a quantum description of the universe, one might start by seeking an appropriate wave function. One potential issue, in trying to realize such a wave function, is that we have to impose a sensible initial condition for the universe. For instance, how long into the past should one place the ``initial'' slicing? A brilliant idea of Hartle and Hawking, called the ``no-boundary proposal'', beautifully evaded this concern by defining the wave function as a sum over geometries with a specific late-time configuration, but ``no-boundary'' in the past. This proposal not only furnished a tangible description of the universe being created from ``nothing'', but also enabled one to fix the initial conditions for perturbations. It is worth mentioning that in the ``tunnelling proposal'', which is an alternate but closely related approach, the relevant wave function realizes the picture of a universe that tunnels into existence from ``nothing''. 
	
	Recently, with the advent of the so-called Lorentzian quantum cosmology (LoQC)\cite{Feldbrugge:2017kzv}, a new level of mathematical rigour has been introduced into the path integral realization of the no-boundary proposal. LoQC enables us to systematically evaluate the amplitude for a universe to emerge from zero-size to a given finite size. However, it was found that the perturbations about the saddle point geometry, when quantized, were unstable\cite{Feldbrugge:2017fcc}. Several proposals were put forward to the address this issue\cite{DiazDorronsoro:2017hti,DiazDorronsoro:2018wro,Vilenkin:2018dch,Vilenkin:2018oja}. However, it was soon argued that some of these proposals either did not entirely resolve the instability or lead to other inconsistencies\cite{Feldbrugge:2017mbc,Feldbrugge:2018gin}. Following this, a reinterpretation of the no-boundary wave function was proposed in \cite{DiTucci:2019dji,Lehners2019}, where the zero-size initial condition was traded off for a certain Robin condition on the scale factor. This latter proposal not only seems to reproduce the Hartle-Hawking wave function but also renders the perturbations stable. Moreover, the saddle point geometry that arises from the path integral also corresponded to the famous Hartle-Hawking geometry. The important distinction, however, was that the off-shell geometries that go into the path integral in \cite{DiTucci:2019dji} were allowed to have arbitrary initial size and momentum, with the average fluctuations around the corresponding mean values akin to that expected from the uncertainty principle. This is in sharp contrast to the original no-boundary proposal, where strictly compact geometries are allowed. Considering the fact that debate over satisfactory implementation and interpretation of the Hartle-Hawking wave function still remains active, what has become even more appreciated is that, despite several important developments, the subject demands to be studied even deeper.       
	
	In the present work, we explore the full quantum cosmological treatment of an inflationary scenario. This is in contrast to the usual approach in inflationary cosmology, where one assumes a classical background spacetime, on top of which quantum fields are studied. It is also important to note that the motivations of the present work are less ambitious, although still quite relevant, than the pioneering works in \cite{Hartle:1983ai} and \cite{Vilenkin:1982de}, where the authors attempted to model the very origin of the Universe. Here, we seek to make progress in studying inflationary scenarios without a priori assuming a fixed background spacetime. To this end, we start off with a novel prescription for the possible quantum state of the universe. The key ingredient of our proposal is a remarkably simple assumption, which also turns out to be the most natural one to make in the context of inflation, namely, that the eventual fate of the universe is to approach an expanding quasi de Sitter phase. We first show explicitly that one can perform a (Lorentzian) path integral over a class of (Lorentzian) universes that are destined to expand -- quite counter-intuitively -- by imposing an appropriate \textit{initial} condition. The underlying principle that motivates this such a path integral definition for the state of the universe is the following: If inflation is a natural outcome of a more fundamental quantum theory of gravitation and matter, one consequence of it could be that the path integral measure for those configurations of the universe corresponding to inflationary scenarios will be quite dominant. The current work considers a toy model that attempts to realizes such a path integral in a much simpler settings. 	Surprisingly, the resulting path integral amplitude is closely related to the Hartle-Hawking wave function. In contrast to several other existing proposals, however, the saddle point geometry corresponds exactly to the classical de Sitter universe. Moreover, there seem to be neither complex nor Euclidean saddle points. 
	
	Another remarkable observation in the quantum analysis of the background geometry is that the corresponding wave function turns out to be independent of a certain parameter $a_0$, that was required to be introduced to impose the initial condition.	In light of this, we investigate the consequence of postulating this invariance, under change of $a_0$, as a principle. In particular, motivated by a recent theorem due to Kontsevich and Segal, and its extension by Witten to cases with dynamical gravity, we find out the domain in complex $a_0-$plane to which one is ``allowed'' to smoothly vary $a_0$. We find that the values of $a_0$ that correspond to the no-boundary proposal lie on the boundary of this domain. This, in turn, demystifies why the wave function arising from our proposal is similar to the Hartle-Hawking wave function. Therefore, the fact that one can analytically continue our result to that of no-boundary prescription leads to a precise sense in which the Hartle-Hawking vacuum is, in fact, just a complex extension of a path integral over universes destined to expand. We also show how one retains the scale-invariant power spectrum of primordial perturbations, using our prescription.

	The structure of the paper is as follows: In \ref{sec_review}, we review the classical dynamics of de Sitter universe, from a viewpoint that is appropriate to the quantum treatment that is to follow. In particular, we motivate the explicit form of the initial condition that unambiguously filters universes that are destined to expand. Then, in \ref{sec_wave_func}, we analyse the quantum aspects of our boundary condition by computing the corresponding wave function from a Lorentzian minisuperspace path integral. Following that, in \ref{sec_Allowable}, we define a class of allowable deformation of the path integral which maintains the physics intact. We explicitly show that one such allowable deformation leads to the no-boundary proposal. We also study the quantum treatment of perturbations in \ref{subsec_NB} and show that our proposal leads to a scale-invariant power spectrum. If one also augment our approach with a version of the no-boundary prescription, we can also retain the Bunch-Davies vacuum. Finally, we conclude by discussing the implications and future direction in \ref{sec_discuss}.

	\section{Classical analysis: a brief revisit}\label{sec_review}
	
	Before going into the quantum cosmological calculations, let us briefly review some classical aspects of the de Sitter spacetime. We shall do this by keeping in mind the particular approach to the quantum problem that we shall deal with in the next section. In particular, we ask ---  classically, how does one impose an \textit{initial} condition that will filter out solutions to Friedman equation that correspond to ``\underline{u}niverses that are \underline{d}estined to \underline{e}xpand'' (UDE)? 
	
	Consider the minisuperspace of closed FRLW spacetimes, with the metric given by:
	\begin{align}
	ds^2=-d\tau^2+a^2(\tau)d\Omega_3^2,
	\end{align}	
	where, $a(\tau)$ is the scale factor and $d\Omega_3^2$ the metric on a unit 3-sphere. When there are no matter fields, and there is a positive cosmological constant $\Lambda \equiv3H^2$, the Friedman equations take the form\cite{Friedman:1922kd}:
	\begin{align}\label{friedman_eq_a_1}
	a''(\tau)-H^2a(\tau)&=0,\\\label{friedman_eq_a_2}
	a'(\tau)^2+1-H^2a(\tau)^2&=0.
	\end{align}
	The first equation above describes the dynamics, while the second is a constraint equation. 
	
\begin{figure}[h!]
		\centering
		\includegraphics[scale=.25]{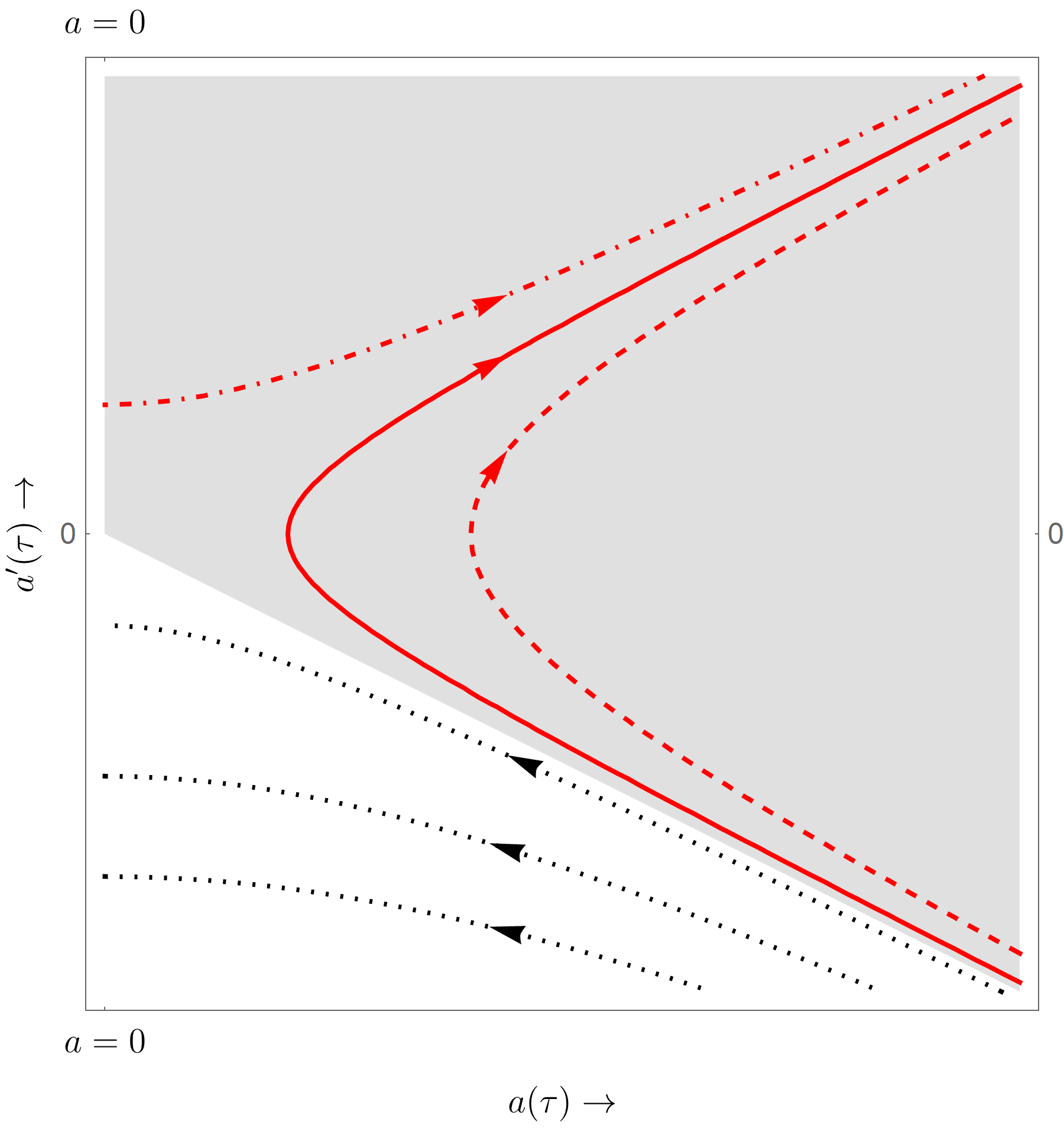}
		\caption{\textbf{Phase space of the scale factor:} The shaded region represents $a'(\tau)+Ha(\tau)>0$ and the curves are different solutions to \ref{friedman_eq_a_1}. Arrows denote the direction of time. The red continuous curve represents the dS universe. The red dashed and red dot-dashed curves represent other solutions to \ref{friedman_eq_a_1} that ends up in an expanding phase. Note that the solutions outside shaded region, represented by black dotted curves, have no expanding phase and eventually contract to singularity.}
		\label{phase_space_1}
	\end{figure}

	Let us now consider the boundary value problem of only \ref{friedman_eq_a_1}; we shall return to the constraint equation shortly. Our aim is to formulate the boundary conditions such that the solution describes a universe which (i) has the size $a_1$ at the final time ( say, $\tau=\tau_1$) and (ii) has the fate of ultimately reaching an expanding phase. Condition (i) is easily implemented as follows:
	\begin{align}\label{final_cond_a}
	a(\tau_1)=a_1.
	\end{align} 
	Naively, one would expect that (ii) must also be implemented as a final boundary condition. However, it turns out that there is a well defined initial condition that \textit{unambiguously} effects (ii), namely,
	\begin{align}\label{initial_cond_a}
	a'(\tau_0)+H a(\tau_0)=\sigma>0\quad;\quad\Rightarrow\sigma\equiv a_0He^{H\tau_0}, 
	\end{align}
	where the positive constant $a_0$ can be fixed to otherwise arbitrary value\footnote{A similar initial condition has already been used in a completely different context in \cite{PhysRevD.104.105014}, Also, see \cite{Rajeev:2021yyl}.}. As we shall see shortly, the specific value of $a_0$ is not quite important, apart from the fact that it is strictly positive. To appreciate how the above initial condition is, in fact, the desired one, first consider the most general solution of \ref{friedman_eq_a_1}, which can be written as $a(\tau)=A_{+}e^{H\tau}+A_{-}e^{-H\tau}$. In the restricted phase space with $a>0$, a solution that reaches an inflating phase in the far future must have $A_{+}>0$. Since $a'(\tau)+H a(\tau)=A_{+}e^{H\tau}$, we find that \ref{initial_cond_a} is nothing but a local version of the condition $A_{+}>0$. To further illuminate this initial condition, it is also instructive to consider the solutions of \ref{friedman_eq_a_1} in the phase space, as presented in \ref{phase_space_1}. The shaded region corresponds to $a'(\tau)+Ha(\tau)>0$ and the curves represent different types of solutions to \ref{friedman_eq_a_1}. We can classify the solutions into three types --- those corresponding to (1) an ever contracting (2) bouncing and (3) ever-expanding universes, respectively. Notice that while types (2) and (3) are inside the shaded region, the solutions of type (1), which does not have an expanding phase, lie outside. This clearly justifies the choice of the initial condition in \ref{initial_cond_a}. It is worth mentioning that this unconventional initial condition is the key ingredient of our proposal, which we also shall shortly incorporate into the path integral formalism and construct the corresponding wave function. However, note that it is not clear whether condition (ii) can be implemented by a simple initial condition analogous to \ref{initial_cond_a} when considering more complicated cosmologies. Moreover, even in the simple case under consideration, the condition \ref{initial_cond_a}, when applied to the off-shell geometries, does not necessarily correspond to the condition (ii). The generalization of our proposal to wider class of cosmological models and implications to off-shell geometries are worth investigating, and shall be topics of investigation in a future work. 
	
	Now that we have made sense of the initial condition, let us write down the corresponding solution to \ref{friedman_eq_a_1}:
	\begin{align}
	a(\tau)&=\frac{1}{2}\left(2a_1-a_0e^{H\tau_1}\right)e^{-H\tau}+\frac{1}{2}a_0e^{H\tau}.
	\end{align} 
	Note that the above solution seems to be independent of $\tau_0$. This indicates that the specific value of the initial time is immaterial, in this boundary value problem. In view of this, although formally \ref{initial_cond_a} is posited as an initial condition, it is reasonable to interpret the same as just \textit{the condition} that filters UDEs. Incidentally, this is a desired feature for the initial condition in quantum cosmology, since one does not have to be concerned with the details of where the initial slice is located. Further, if we impose the constraint \ref{friedman_eq_a_2} on the above solution, we arrive at the final result:
	\begin{align}
	a(\tau)=\frac{1}{H}\cosh\left[H\tau-c_0\right]\quad;\quad c_0\equiv-\log(a_0H),
	\end{align}
	This is nothing but the standard scale factor for the dS universe in closed slicing\cite{de1917curvature}, as expected. It is clear from the above equation that $a_0$ determines the location of the ``throat'' of dS. Alternatively, one can also imagine $a_0$ as a parameter that determines the location of the ``initial'' space-like slice. Therefore, at first glance, it seems that the introduction $a_0$ re-invokes the issue of fixing the initial slicing, which we thought we had already evaded. However, in the next section, we will show that the appropriate wave function is independent of $a_0$, apart from an overall phase. Hence, in a sense, $a_0$ is merely a dummy parameter, whose sole purpose is to ensure that only UDEs are being summed over in the path integral.
	
	\section{Quantum analysis: the background}\label{sec_wave_func}
	We shall now compute the wave function that arises out of our prescription in the last section. In the path integral formalism, the desired wave function $\Psi(a)$ can be formally defined as follows:
	\begin{align}\label{PI_sketch}
	\Psi(a)=\sum_{\textrm{UDEs}}e^{\frac{i}{\hbar}S},
	\end{align}
	where $S$ is the appropriate action and $\sum_{UDEs}$ is short-hand for the path integral over UDEs with the final boundary corresponding to a universe of size (i.e., scale factor) $a$. As is the case with the no-boundary proposal or the tunnelling proposal, it is technically a challenge to compute the amplitude in \ref{PI_sketch}. Hence, to make progress, we shall work in the minisuperspace, where the metric is conveniently parametrized in the following form\cite{Halliwell:1988ik}:
	\begin{align}
	ds^2=-\frac{N(t)^2}{q(t)} dt^2+q(t)d\Omega_3^2,
	\end{align}  
	where $q\equiv a^2$, the time coordinate $t$ is defined via $Ndt=a(\tau)d\tau$ and $N(t)$ is the lapse function. We shall henceforth stick to the gauge\footnote{In the path integral formalism, this gauge fixing can be implemented through the BFV quantization \cite{Batalin:1977pb}, as for instance, has been done in \cite{Halliwell1988}.} $\dot{N}=0$ and assume, without lose of generality, that $t=0$ is the ``initial time'' and $t=1$ is the ``final time''. The boundary conditions, \ref{final_cond_a} and \ref{initial_cond_a}, in this parametrization translates to:
	\begin{align}\label{final_cond_q}
	q(1)&=q_1=a_1^2,\\\label{initial_cond_q}
	\mathcal{I}[q(0)]&\equiv\frac{\dot{q}(0)}{N}+2H\sqrt{q(0)}=2H a_0\quad;\quad a_0>0.
	\end{align}
	It is worth mentioning that similar, but different, conditions have been previously employed in the literature to implement the no-boundary proposal\cite{Lehners2019}. However, here \ref{initial_cond_q} (or equivalently \ref{initial_cond_a}) follows as the natural condition to be implemented to realize our proposal of summing over UDEs. 
	
	Now, for convenience, we shall first work with the following Neumann final condition:
	\begin{align}\label{final_cond_p}
	\frac{\dot{q}(1)}{N}=p_1,
	\end{align}
	instead of \ref{final_cond_q}. Although we can keep $p_1$ arbitrary at this point, for problems of interest to late-time cosmology, we might choose its value to be positive and large($p_1\gg 1$). The wave function corresponding to the Dirichlet condition in \ref{final_cond_q} can be easily found by Fourier transforming. Moreover, recall that the Neumann condition is natural to the standard Einstein-Hilbert action. To implement the non-trivial condition \ref{initial_cond_q}, we need to amend the action with appropriate boundary terms. The desired action turns out to be (with $\kappa=8\pi G/c^4$)\cite{Krishnan:2017bte}:
	\begin{align}
	S= \frac{1}{2\kappa}\int\sqrt{-g}\left(R-6H^2\right)d^4x-\frac{H}{\kappa}\int_{\partial\mathcal{M}_0}\sqrt{h}d^3y,
	\end{align}
	where $\partial M_{0}$ denotes space-like boundary at $t=0$ and $h_{ij}$ denotes the induced metric on $\partial\mathcal{M}_{0}$. In terms of the minisuperspace variable, the action reduces to:
	\begin{align}
	\frac{\kappa S}{3V_3}=\int_{0}^{1}\left[\frac{q\ddot{q}}{2N}+\frac{\dot{q}^2}{4N}-H^2Nq+N\right]dt-\frac{H}{3}q(0)^{3/2},
	\end{align}
	where $V_3$ is the volume of unit 3-sphere. Hence, the desired wave function $\tilde{\Psi}$ can be evaluated as the following path integral. 
	\begin{align}\label{PI_momentum_space}
	\tilde{\Psi}(p_1;a_0)=\int_{-\infty}^{\infty}dN\int_{\mathcal{I}[q(0)]=2Ha_0}^{\dot{q}(1)/N=p_1}\mathcal{D}[q]e^{\frac{i}{\hbar}S[q,N]}.
	\end{align}
	Note that we have also exhibited $a_0$ in the argument of $\tilde{\Psi}$, to account for the potential dependence of the wave function on the same. Also, note that the range of integration of the lapse $N$ is from $-\infty$ to $\infty$, as opposed to the positive real line. The latter choice has been used in the pioneering works of Claudio Bunster (known previously as Claudio Teitelboim)\cite{Teitelboim:1981ua,Teitelboim:1983fh,Teitelboim:1983fk}, where the path integral formalism for studying propagators in quantum gravity was developed. The reason we are, instead, integrating $N$ over the entire real line is that we are interested in evaluating a wave function, i.e., a solution to the corresponding Wheeler-DeWitt equation, as opposed to a propagator, i.e., the Green's function of the corresponding Wheeler-DeWitt operator. In fact, the integration of lapse over entire real line has already been in use in the literature, for instance in \cite{DiTucci:2019bui}, to study solutions of Wheeler-DeWitt equation. However, for completeness, we shall now briefly discuss some important points concerning the interpret of the wave functions $\tilde{\Psi}(p_1;a_0)$, from a path integral perspective.    
	
	To this end, let us first consider a much simpler problem, namely the definite momentum wave function $\psi_{\mathbf{p}}(\mathbf{x})=e^{i\mathbf{p}.\mathbf{x}/\hbar}$ in non-relativistic quantum mechanics. The physical interpretation of $\psi_{\textbf{p}}$ is that it gives the amplitude for the particle to be at a given point $\mathbf{x}$, \textit{given} that it has a definite momentum $\textbf{p}$; the argument of the wave function is $\mathbf{x}$ and the momentum appears as a parameter. If the particle is free, one does not have to specify `when' did the particle have the momentum $\textbf{p}$. Now, consider the following PI realisation $\psi_{\textbf{p}}(x)$:
		\begin{align}
		\psi_{\textbf{p}}(\mathbf{x})=\int_{-\infty}^{\infty}dN\int_{m\dot{\mathbf{q}(0)}/N=\mathbf{p}}^{\mathbf{q}(1)=\mathbf{x}}\mathcal{D}[\mathbf{q}]\exp\left[\frac{i}{\hbar}\int_{0}^{1}\left(\frac{m}{2N}\dot{q}^2+\frac{p^2}{2m}N\right)dt+\frac{i}{\hbar}S_{b}\right];\qquad S_b=\frac{m \dot{\mathbf{q}}(0).\mathbf{q}(0)}{N}.
		\end{align}   
		Note that in the above, one introduces the definite momentum $\mathbf{p}$ through the `initial' condition $m\dot{\mathbf{q}(0)}/N=\mathbf{p}$. However, since we are integrating the `lapse' $N$ over the whole real line, $t=0$ is not strictly an `initial' instance for all the paths contribution to $\psi_{\textbf{p}}(\mathbf{x})$. So a more sensible interpretation of $m\dot{\mathbf{q}(0)}/N=\mathbf{p}$ is that it is \textit{the} condition (i.e., without any adjectives) that imposes the definite momentum. This is merely a manifestation of the fact that our interpretation of $\psi_{\textbf{p}}(\mathbf{x})$ does not necessarily demand `\textit{when}', precisely, must the momentum of the particle be fixed at $\mathbf{p}$. 
	
	Now, let us come back to the interpretation of $\tilde{\Psi}(p_1;a_0)$. Actually, it is more natural to compare the non-relativistic wave function $\psi_{\textbf{p}}(\mathbf{x})$ with $\Psi(q_1;a_0)$, which is just a Fourier transform of $\tilde{\Psi}(p_1;a_0)$. $\Psi(q_1;a_0)$ is to be interpreted as a wave function of just one argument, namely $q_1$(size of the universe), while $a_0$ appears as a parameter. However, in the PI realisation of the same wave function, one introduces this parameter through what seems like an `initial' condition, say as in \ref{PI_momentum_space}. But, just as in the case of definite momentum state, a more sensible interpretation of the apparent `initial' condition is that it is \textit{the} condition that imposes the UDE prescription. Therefore, the interpretation of $\Psi(q_1;a_0)$ is that it is the amplitude for the universe to have a given size (namely, the one corresponding to $q_1$), given that the UDE condition is imposed (i.e., analogous to the definite momentum condition in the case of $\psi_{\mathbf{p}}(\mathbf{x})$). In light of \ref{initial_cond_q} one might be tempted to take the analogy further and identify $\Psi(q_1;a_0)$ as the eigenvector of the operator $\hat{O}\equiv-\hat{P}_{q}/H+\sqrt{\hat{q}}$ (with the corresponding eigenvalue being $a_0$), where $P_{q}$ is the momentum conjugate to $q$. But, it is easily verified that this is not the case. However, in terms of the actual eigenfunction $\xi_{a_0}(q_1)$ of the operator $\hat{O}$ we can rewrite (the Fourier transform of) \ref{PI_momentum_space} as
		\begin{align}
		\Psi(q_1;a_0)=\int_{-\infty}^{\infty} dN\int_{-\infty}^{\infty}dq_2 G(q_1,q_2;N)\xi_{a_0}(q_2)dq_2=\int_{-\infty}^{\infty}dN\braket{q_1|e^{-i \mathcal{H} N}|\xi_{a_0}}
		\end{align}
		where, $\braket{q_1|e^{-i \mathcal{H} N}|q_2}\equiv G(q_1,q_2; N)$ and $\mathcal{H}N$ is the Hamiltonian.  Hence,$\Psi(q_1;a_0)$ is related to the transition amplitude from a fixed $\hat{O}$ state to a fixed size state. Incidentally, this furnishes an interpretation of the parameter $a_0$ that goes beyond the semi-classical level.
	
\begin{figure}[h!]
		\centering
		\begin{subfigure}[b]{0.48\textwidth}
			\centering
			\includegraphics[width=\textwidth]{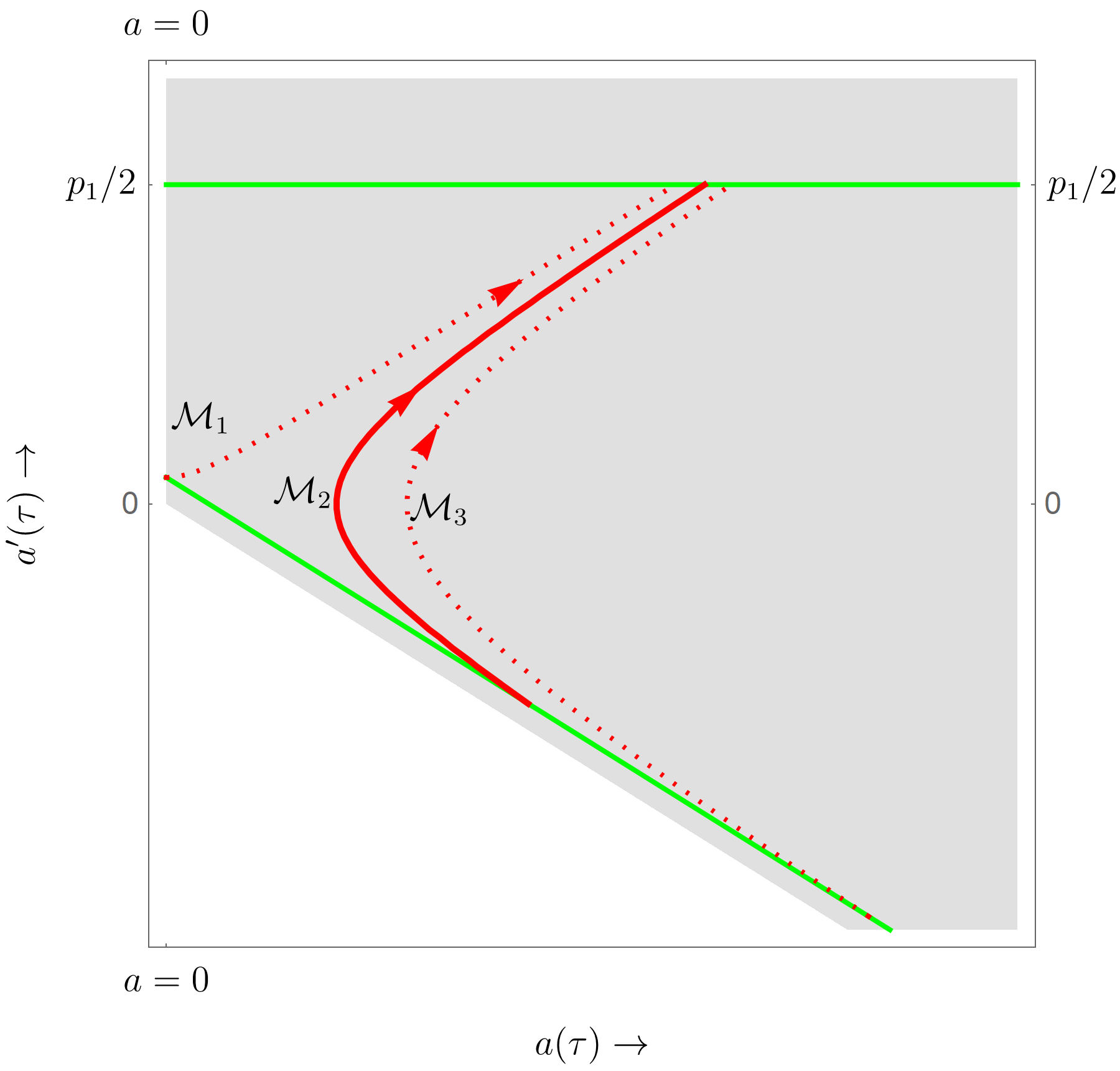}
			\caption{}
			\label{phase_space_2}
		\end{subfigure}
		\begin{subfigure}[b]{0.48\textwidth}
			\centering
			\includegraphics[width=\textwidth]{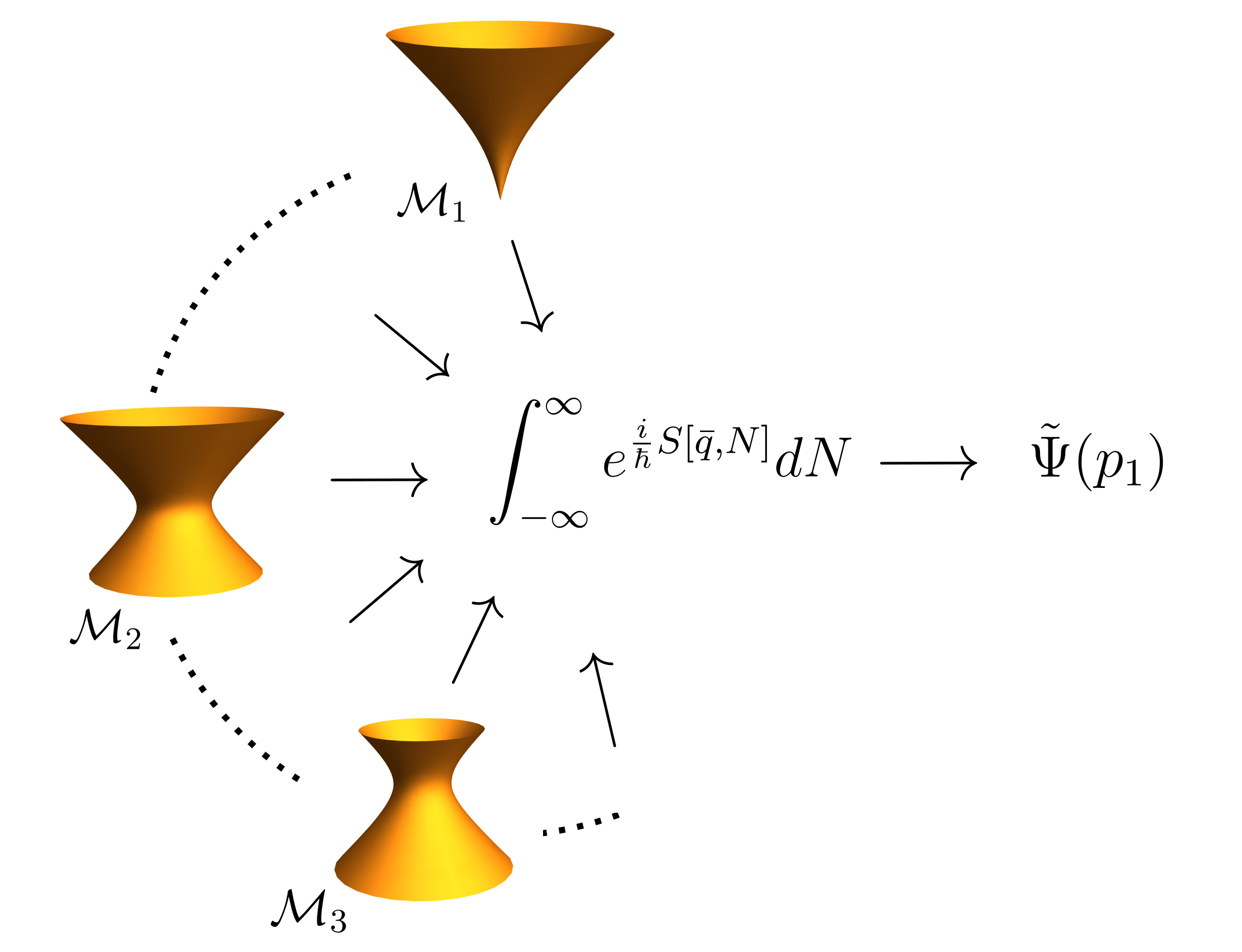}
			\caption{}
			\label{sum_manifold}
		\end{subfigure}
		\caption{\textbf{Description of $\tilde{\Psi}(p_1;a_0)$ as sum over UDEs:} (a) The horizontal green line denotes the boundary condition $\dot{q}(1)/N=2a'(\tau_1)=p_1$ and the slopped green line represents $\dot{q}(0)/N+2H\sqrt{q(0)}=2H a_0$. The other curves represent three types of universes which are consistent with our boundary conditions and contribute to the integral in \ref{Lapse_int} (see text for more details). (b) A cartoon representation of the sum over manifolds that leads to the wave function $\tilde{\Psi}(p_1;a_0)$.  }
		\label{sum_manifold_full}
	\end{figure}

	Now, we return to the computation of $\tilde{\Psi}(p_1;a_0)$.
	Since the action is quadratic in $q$, upto a prefactor, the path integral over $q$ simplifies to\footnote{We shall not focus on the pre-factor here, since we are mainly interested in the semi-classical limit of the wave function.}:
	\begin{align}\label{Lapse_int}
	\tilde{\Psi}(p_1;a_0)\approx\int_{-\infty}^{\infty}e^{\frac{i}{\hbar}S[\bar{q},N]}dN.
	\end{align}
	Here, $\bar{q}$ denotes the classical solution consistent with the boundary conditions given in \ref{final_cond_p} and \ref{initial_cond_q}. The explicit form of $\bar{q}(t)$ is found to be:
	\begin{align}
	\bar{q}(t)&=H^2N^2t^2+Nt\left(p_1-2H^2N\right)+\frac{\left[p_1-2 H (a_0+H N)\right]^2}{4 H^2},
	\end{align} 
	and the corresponding action evaluates to:
	\begin{align}\label{on_shell_action}
	\frac{\kappa S[\bar{q};N]}{3V_3}=-a_0H^3 N^2+N \left[a_0 H (p_1-a_0 H)+1\right]-\frac{(2 a_0 H-p_1)^3}{24 H^2}
	\end{align} 
	Before we move on, let us appreciate the implication of \ref{Lapse_int}. The integral can be clearly interpreted as a sum over different \textit{Lorentzian} universes, each corresponding to a different $N$, but all of them satisfying the desired boundary conditions. We can conveniently classify these universes into three types --- (1) those emerging from the singularity ($\mathcal{M}_1$), reminiscent of those relevant to the no-boundary proposal, (2) the de Sitter universe ($\mathcal{M}_2$) and (3) other bouncing universes ($\mathcal{M}_3$), which are \text{not} de Sitter. The value of lapse for these three types of universes correspond to, respectively, $N<N_{s}$, $N=N_{s}$ and $N>N_{s}$, where $N_{s}$, whose explicit form we shall present shortly, is the saddle point of the action $S[\bar{q},N]$. We have presented one representative spacetime from each of these types in \ref{sum_manifold_full}. There, in \ref{phase_space_2}, the horizontal green line represents the initial boundary condition \ref{final_cond_p} and the slopped green line denotes the condition in \ref{initial_cond_q}. The thick red curve is the de Sitter universe, while the red dotted curves are other UDEs that contribute to the wave function $\tilde{\Psi}(p_1;a_0)$. In \ref{sum_manifold}, we have also provided a cartoon representation of the information in \ref{phase_space_2}. These figures clearly illustrate the precise sense in which $\tilde{\Psi}(p_1;a_0)$ is a sum over UDEs. 
	
	The saddle-point of the classical action $S[\bar{q},N]$, defined via $\partial_{N} S=0$, turns out to be:
	\begin{align}
	N_{s}(a_0)=\frac{a_0 H (p_1-a_0 H)+1}{2 a_0 H^3},
	\end{align}
	and the corresponding on-shell solution yields:
	\begin{align}\label{onshell_q}
	\bar{q}_{s}(t)&=\frac{1}{H^2}+\frac{\left[a_0 H (p_1-a_0 H)+1\right]^2 }{4 a_0^2
		H^4}\left[t-\frac{a_0^2 H^2-1}{a_0 H (a_0 H-p_1)-1}\right]^2.
	\end{align}
	From the above equation, it is clear that the sizes of the spatial slices of the on-shell geometry do not vanish for any real value of $t$. Therefore, in particular, the on-shell solution does not satisfy the no-boundary condition. In fact, it is easily verified that $\bar{q}_{s}$, written in terms of the cosmic time $\tau$, gives rise to the standard scale factor for the de Sitter universe. That is:
	\begin{align}
	a_{s}(\tau)\equiv\sqrt{\bar{q}_s}=\frac{1}{H}\cosh(H\tau),
	\end{align}
	which indicates that $\bar{q}_{s}$ describes pure de Sitter spacetime. The saddle point approximation of $\tilde{\Psi}$, apart from an overall normalization, can be evaluated to get:
	\begin{align}
	\tilde{\Psi}(p_1;a_0)&\approx e^{-\frac{i}{\hbar}S[\bar{q}_{s},N_s]}\\
	&=e^{-i\phi(a_0)}e^{\frac{iV_3}{H^2\ell_p^2}\left( \frac{p_1^3}{24}+\frac{ p_1 }{2}\right)}
	\end{align}
	where we have defined the Planck-length($\ell_p$) through $3\kappa/\hbar=\ell^2_p$ and the phase is given by $\phi(a_0)=V_3(a_0^4H^4+6a_0^2H^2-3)/(4a_0H^3\kappa\hbar)$. Note that the only dependence of the wave function on the parameter $a_0$ is through an overall phase, which may be understood as a factor that is required to ensure $\tilde{\Psi}(p_1;a_0)$ solves the WdW equation at $t=0$ as well. This implies that, as far as physically relevant computations at $t=1$ are concerned, the wave function $\tilde{\Psi}$ may be regarded as independent of $a_0$. This ``symmetry'', as we shall shortly see, leads to interesting consequences. 
	
	Now, a simple Fourier transform gives the following wave function $\Psi(q_1;a_0)$ appropriate to the final Dirichlet condition in \ref{final_cond_p}.
	\begin{align}\label{def_psi_q1}
	\Psi(q_1;a_0)\equiv \int_{-\infty}^{\infty}\tilde{\Psi}(p_1;a_0)e^{-\frac{i}{2}\frac{V_3}{\ell^2_p}p_1q_1}dp_1\propto e^{-i\phi(a_0)}\textrm{Ai}\left[\frac{V_3^{2/3}(1-H^2q_1)}{(H\ell_p)^{4/3}}\right]
	\end{align}
	Note that the $\Psi(q_1)$ is \textit{proportional} to the Hartle-Hawking wave function. For instance, the semi classical limit (i.e., $H\ell_p\ll1$) gives the following:
	\begin{align}
	\Psi(q_1;a_0)\approx \frac{e^{-i\phi(a_0)}}{\sqrt{H^2q_1-1}}\cos\left[\frac{2}{3}\frac{V_3 }{(H\ell_p)^2}\left(q_1H^2-1\right)^{3/2}-\frac{\pi}{4}\right]\quad;\quad q_1H^2>1\textrm{ and }H\ell_p\ll 1,
	\end{align} 
	which may be compared with the well known asymptotic behaviour of the Hartle-Hawking wave function\cite{Hartle:1983ai}. Therefore, we seem to have obtained some features of the no-boundary wave function without invoking any assumptions that directly or indirectly implement the no-boundary proposal. Moreover, when $q_1<1/H^2$, the $p_1$ integration in \ref{def_psi_q1} is dominated by the saddle point value $p_s=i\sqrt{1-H^2q_1}$, implying that the corresponding geometry consists of Euclidean sections. Using this, one finds that the wave function takes the form
		\begin{align}
			\Psi(q_1;a_0)\propto \frac{e^{-i\phi(a_0)}}{\sqrt{1-H^1q_1}}e^{-\frac{2}{3}\frac{V_3}{(H\ell_p^2)}(1-H^2q_1)^{3/2}}
		\end{align}   
The above structure of $\Psi(q_1;a_0)$ reveals that it also coincides with the amplitude for the Universe to emerge from zero-size.

	What is even more surprising is that the natural state of the perturbations, that arise out of our prescription, turns out to retain the scale-invariant spectrum! Before going into that analysis, we shall next explore, in greater detail, the ``symmetry'' of the wave function under the change of parameter $a_0$.

	\section{Allowable analytical continuations of the path integral}\label{sec_Allowable}
	
	Recall that our wave function $\Psi(q_1;a_0)$ (or equivalently $\tilde{\Psi}(p_1;a_0)$) was independent of the parameter $a_0$, modulo an overall phase factor given by $e^{-i\phi(a_0)}$. Hence, as long as $a_0>0$, it is reasonable to think of \ref{PI_momentum_space} as representing path integral over UDEs, with $a_0$ introduced merely as a place-holder to enable a convenient implementation of the corresponding boundary condition. In view of this, it is worth exploring whether one can smoothly vary parameter $a_0$ to an extended domain in the complex $a_0$-plane while retaining the same \textit{physical content} as the original wave function.  For this purpose, let us first write an absolutely convergent integral representation\footnote{It is important that, before performing analytic continuation, we start with a convergent integral representation for $\tilde{\Psi}(p_1;a_0)$. Because, Fubini's theorem\cite{zbMATH02643959} would then allow one to perform path integrals over different degrees of freedom in any order without affecting the final result. } for $\tilde{\Psi}(p_1;a_0)$, by deforming the real-line contour in \ref{Lapse_int} to $\mathcal{J}$, the union of corresponding Lefschetz thimbles:
	\begin{align}\label{Lapse_int_thibles}
	\tilde{\Psi}(p_1;a_0)\approx\int_{\mathcal{J}}e^{\frac{i}{\hbar}S[\bar{q},N]}dN.
	\end{align}
	The contour $\mathcal{J}$ turns out to be a straight line with slope $=-1$, and passing through the saddle point $N_s$. We can now define the ``allowable domain'' $\mathcal{D}$ of $a_0$ as the region in the complex $a_0-$plane, to which one can perform the smooth analytical continuation $a_0\rightarrow a_{r}+i a_{i}$, from $a_0\in\mathbb{R}^{+}$,  while still maintaining the above integral well defined. It is easily seen that the allowable domain is given by $\mathcal{D}=\{a_0\,\,|-\frac{\pi}{2}<\textrm{arg}(a_0)<\frac{\pi}{2}\}$.

	Two interesting special cases arise in the boundary of $\mathcal{D}$, namely, $a_0\rightarrow H^{-1}e^{\pm i\pi/2}$. The corresponding on-shell solutions turn out to be:
	\begin{align}\label{onshell_nb_q}
	\bar{q}_{\pm}(t)=\frac{(p_1\mp2 i) }{4 H^2}t \left[(p_1\mp 2 i) t\pm 4 i\right]
	\end{align} 
	Clearly, $\bar{q}_{\pm}(0)=0$. This means that the saddle point relevant to the analytic continuation of the path integral in \ref{Lapse_int}, corresponding to $a_0\rightarrow H^{-1}e^{\pm i\pi/2}$, coincides with that of the no-boundary proposal. This result also demystifies why the wave function that arose from our prescription is similar to the Hartle-Hawking wave function. However, there is a subtle and important difference between the continuations $a_0\rightarrow H^{-1}e^{\pm i\pi/2}$ --- the pre-factor, arising from analytic continuation of the phase $\psi(a_0)$, turns out to be $\exp\left[2V_3/(H^2\kappa\hbar)\right]$ for the former, while it is $\exp\left[-2V_3/(H^2\kappa\hbar)\right]$ for the latter. Incidentally, $a_0\rightarrow H^{-1}e^{i\pi/2}$ furnishes a well-defined Euclidean path integral realization of the no-boundary wave function, contrary to several recent investigations that seems to indicate that Euclidean integrals lead to divergent results. More about this will be discussed in \ref{subsec_NB}.
	
	\subsection{Kontsevich-Segal theorem}
	We have so far assumed that apart from the convergence of \ref{Lapse_int_thibles} that defines $\tilde{\Psi}(p_1)$, there are no other physical principles that restrict us from smoothly continuing the positive parameter $a_0$ to the complex domain $\mathcal{D}$. It is not at all clear whether this is a reasonable assumption. In fact, we shall now present an argument that seems to suggest that there are potentially more restrictions on the manner in which $a_0$ can be varied. To this end, we shall consider an important theorem, proposed recently by Kontsevich and Segal\cite{Kontsevich:2021dmb}, which says that, in a certain precise sense, a large class of physically interesting quantum field theories can be sensibly realised on manifolds endowed with complex metric $\textbf{g}(x^{\mu})$, if its eigenvalues $\{\lambda_i(x^{\mu})\}$ satisfy:
	\begin{align}\label{KS}
	\sum \left|\textrm{arg}\left(\lambda_i\right)\right|<\pi
	\end{align}
	The Kontsevich-Segal(KS) theorem essentially afford us a refined notion of physically ``allowable'' complex metrics. Witten has recently investigated several implications of extending the KS theorem to contexts where gravity is dynamical \cite{Witten:2021nzp}. Lehners has also extended this analysis for off-shell geometries of certain minisuperspace gravitational path integrals\cite{Lehners:2021mah}. Here, we seek the region in complex $a_0$-plane for which the on-shell solution $\bar{q}_{s}(t)$ leads to allowable metrics a la KS. This amounts to finding the region in which
	\begin{align}\label{KS_specific}
	\Theta(a_0)\equiv \left|\textrm{arg}\left(\frac{-N_s^2}{\bar{q}_s(t)}\right)\right|+3\left|\textrm{arg}\left(\bar{q}_{s}(t)\right)\right|<\pi
	\end{align}
	In a sense, what we are demanding via the above condition is that along with the path integral for gravitation, that for a wide class of matter field should also remain convergent as $a_0$ is varied. 
	
	For real values of $a_0$, the bound in \ref{KS_specific} is easily seen to saturate. To investigate the case of a complex $a_0$, we begin by noting that the ``physical time'' given by $T\equiv \int_{0}^{1}(N_s/\sqrt{\bar{q}_s})dt$ is complex valued, which in turn means that the evolution possibly takes place along interchanging time signatures. Now, for convenience, one can make a gauge transformation from $N=N_s$ to $N\rightarrow N(t)$ such that near $t\approx 0$ the signature is approximately Euclidean and near $t\approx 1$ the same is Lorentzian, while keeping the total elapsed (complex) physical time $T$ unchanged. The advantage of this gauge choice is the following. Note that, in the new gauge, the first term of $\Theta(a_0)$ in \ref{KS_specific} is negligible near $t\approx 0$. On the other hand, near $t\approx 1$, while the first term of $\Theta(a_0)$ will be slightly less that $\pi$, the second term vanishes, owing to $\bar{q}_{s}(1)=(4+p_1^2)/(4H^2)\in \mathbb{R}$. Therefore, the factor that essentially determines the satisfaction of \ref{KS_specific} turns out to be the second term near $t\approx 0$. More specifically, the allowable region is well described by $3|\textrm{arg}(\bar{q}_s(0))|<\pi$. 
	
\begin{figure}[h!]
		\centering
		\includegraphics[scale=.2]{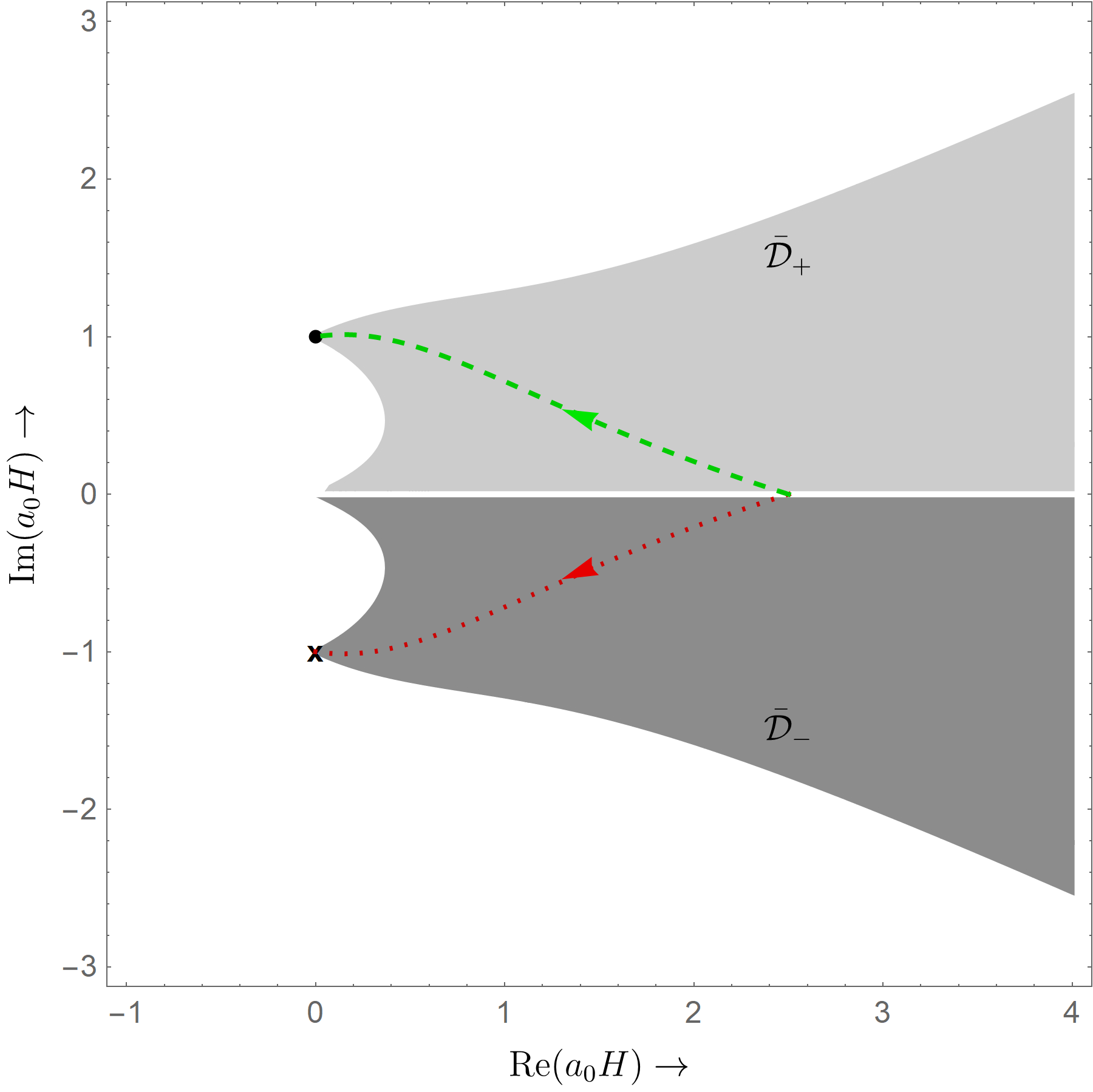}
		\caption{\textbf{Allowable continuations of $a_0$}: The shaded region represents the subset of $\mathcal{D}$ in which $3|\textrm{arg}(\bar{q}_s(0))|\leq\pi$. The black dot and cross are, respectively, the points $Ha_0=i$ and $Ha_0=-i$. Two representative allowable paths of analytic continuation of $a_0$ from the positive real line to $a_0=\pm i/H$ are represented by the green dashed and red dotted curves, respectively.}
		\label{KS_plot}
	\end{figure}

	The intersection of the region satisfying $3|\textrm{arg}(\bar{q}_s(0))|<\pi$ and $\mathcal{D}$, which we shall refer to as $\bar{\mathcal{D}}$, is represented by the shaded region in \ref{KS_plot}. Note that $\bar{\mathcal{D}}$ consists of two disconnected regions, represented by lighter ($\bar{\mathcal{D}}_{+}$) and darker ($\bar{\mathcal{D}}_{-}$) shaded regions in \ref{KS_plot}, which are separated by the positive real line $\mathbb{R}^{+}$. Since $\mathbb{R}^{+}$ itself is on the boundary of both $\bar{\mathcal{D}}_{\pm}$, one should view the positive real line as one of the two inequivalent limits, namely, of approaching $\textrm{Im}[a_0]=0$ from either $\bar{\mathcal{D}}_{+}$ or $\bar{\mathcal{D}}_{-}$, respectively. The black dot and cross in \ref{KS_plot} are, respectively, the points $Ha_0=-i$ and $Ha_0=i$. Not that one can smoothly vary $a_0$ from the positive real line to $a_0=\pm i/H$ without violating the KS criterion. The green dashed curve and red dotted curve in \ref{KS_plot} are possible allowed paths leading to $a_0=\pm i/H$, respectively.  
	
	We shall now explore the consequences of (1) the sum over UDEs and (2) the invariance under the smooth variation of $a_0$, for the analysis of primordial perturbations.
	
	\section{Quantum analysis: the perturbations}\label{subsec_NB}

	The second-order action for a given mode of gravitational perturbation, labelled by $l$, is given by
	\begin{align}\label{mat_action}
	S^{(2)}_{l}[\chi,q,N]=\frac{V_3}{2}\int_{0}^{1} dt N\left[q^2\frac{\dot{\chi}^2}{N^2}-l(l+2)\chi^2\right]
	\end{align}
	Therefore, one can define the semi classical limit of the wave function $\tilde{\Psi}_{tot}(p_1,\chi_1)$, for the background and perturbations combined, as follows:
	\begin{align}\label{mat_PI}
	\tilde{\Psi}_{tot}(p_1,\chi_1)\approx e^{-\frac{i}{\hbar}S[\bar{q}_{s},N_s]}\int_{-\infty}^{\infty}d\chi_0\int^{\chi(1)=\chi_1}_{\chi(0)=\chi_0}\mathcal{D}[\chi]e^{iS^{(2)}_{l}[\chi,\bar{q}_s,N_s]}\psi_{0}(\chi_0)
	\end{align} 
	where, for generality, we have introduced an ``initial state'' $\psi_{0}$ for the perturbation. If one is interested in the lowest excitations of the system, it is reasonable to consider a Gaussian form for $\psi_{0}$. In this spacial case, the $\chi-$dependent part of \ref{mat_PI} describes the unitary evolution of a Gaussian state from $t=0$ to $t=1$. Note that the action $S^{(2)}_{l}[\chi,\bar{q}_s,N_s]$ in \ref{mat_action} describes a time dependent harmonic oscillator (TDHO) of frequency $\omega_{l}(t)$ and mass $m(t)$ given by:
	\begin{align}
	m(t)=\frac{V_3}{N_s}\bar{q}_s^2(t)\qquad;\qquad \omega^2_l(t)=\frac{N_s^2l(l+2)}{\bar{q}^2_s(t)}
	\end{align}
	Therefore, time evolution of a Gaussian state can be written as:
	\begin{align}\label{evolve_gaussian}
	\psi(\chi_1^2,t)=\mathcal{N}_l(t)\exp\left[i\frac{m}{2}\frac{\dot{f}_l(t)}{f_l(t)}\chi_1^2\right]
	\end{align}
	where $\mathcal{N}_l$ is a normalization factor and $f_l(t)$ is a suitable solution of the following TDHO equation:
	\begin{align}\label{TDHO}
	\ddot{f}_l+\frac{\dot{m}}{m}\dot{f}_l+\omega^2_{l}f_l=0
	\end{align}
	The two linearly independent solutions of the above equation may be chosen as:
	\begin{align}\label{pert_solns}
	\chi_{\pm}(t)=\frac{1}{\sqrt{\bar{q}_s(t)}}\left[\frac{t-\bar{\gamma}(a_0)}{t-\gamma(a_0)}\right]^{\pm\frac{(l+1)}{2}}\left\{[1\mp(l+1)][\gamma(a_0)-\bar{\gamma}(a_0)]+2[t-\gamma(a_0)]\right\}
	\end{align}
	where $\{\gamma(a_0),\bar{\gamma}(a_0)\}$ are the two roots of the equation $\bar{q}_{s}(t)=0$, with
	\begin{align}
	\gamma(a_0)=\frac{\left(a_0 H+i\right)^2}{a_0 H \left(a_0 H-p_1\right)-1}\qquad;\qquad\bar{\gamma}(a_0)=\frac{\left(a_0 H-i\right)^2}{a_0 H \left(a_0 H-p_1\right)-1}
	\end{align}
	
	Given a general solution $f_l(t)=\chi_{+}(t)+\beta\,\,\chi_{-}(t)$ of \ref{TDHO}, with the constant $\beta\in\mathbb{C}$, one can construct a time evolving Gaussian state using \ref{evolve_gaussian}, as long as the initial conditions on $f_{l}(t)$ are such that the wave function is normalizable. Consequently, it seems that the best we can progress from \ref{mat_PI} is to write down a family of wave functions of the form:
	\begin{align}
	\tilde{\Psi}_{tot}(p_1,\chi_1)\approx e^{-\frac{i}{\hbar}S[\bar{q}_{s},N_s]}\exp\left[\frac{i}{\hbar}\tilde{S}[\chi_1]\right]
	\end{align} 
	where, 
	\begin{align}
	i\tilde{S}[\chi_1]&=i\frac{m(1)}{2}\frac{\dot{f}_l(1)}{f_l(1)}\chi_1^2\\\label{action_pert}
	&=-\frac{i l (l+2) p_1 V_3
		\chi _1^2}{4 H^2}+\frac{(\beta -1) l (l+1) (l+2) V_3 \chi _1^2}{2 (\beta +1) H^2}+\mathcal{O}(p_1^{-1})
	\end{align}
	Note that, interestingly, the total wave function $\tilde{\Psi}_{tot}(p_1,\chi_1)$ is also independent of $a_0$, apart from a constant phase. From \ref{action_pert}, we conclude that if $|\beta|^2<1$, the perturbations are described by a Gaussian state that gives rise to a scale-invariant power spectrum. Now, if we also assume, motivated by the symmetry of the background wave function $\tilde{\Psi}(p_1)$, that the total wave function satisfies $\tilde{\Psi}_{tot}(-p_1,\chi_1)\propto \tilde{\Psi}^*_{tot}(p_1,\chi_1)$, then we get the further constraint $\beta\in\mathcal{R}$, so that we finally have $-1<\beta<1$. However, it appears that we cannot argue any further and converge on a ``unique'' value for the parameter $\beta$. Consequently, for instance, we do not seem to arrive at the Bunch-Davies state uniquely, but instead, we obtain a family of states, of which the Bunch-Davies state happens to be a special case, namely, when $\beta=0$. 
	
	It was recently argued in \cite{DiTucci:2019xcr} that the inflationary paradigm is not ``quantum complete'', in the sense that the model by itself cannot explain the origin of primordial perturbations.  Our study appears to complement this result since we have also found that the natural path integral boundary condition for the inflationary model, namely the sum over UDEs, do not by itself uniquely fix the state of primordial perturbations. However, we emphasize that the sum over UDEs gives rise to a family of scale-invariant power spectra, with one undetermined parameter $\beta$. Therefore, it is fair to say that the inflationary hypothesis very well explains important \textit{qualitative} quantum aspects of the primordial perturbations. Next, we shall see how, by augmenting the ``sum-over-UDEs approach''  with an additional principle motivated by the no-boundary proposal, we can actually fix the parameter $\beta$.  
	
	\subsection{An avatar of the no-boundary proposal}

	As we have alluded to before, the allowable analytic continuation $a_0\rightarrow i/H$ is of special interest, because it gives the no-boundary wave function with the correct pre-factor. Moreover, it can be shown that the steepest descent contour for evaluating the lapse integral, for this continuation, is along the purely Euclidean direction in the complex $N-$plane. We have already found that the corresponding saddle point describes a universe evolving from zero size. \footnote{Note that although $a_0\rightarrow i/H$ does, in fact, give the no-boundary wave function, the defining path integral of the same does not strictly correspond to the original no-boundary proposal. The reason being, although the saddle point geometry in this case satisfies $\bar{q}_{+}(0)=0$, there are contributions from off-shell geometries which, while significantly contributes to the wave function, necessarily violates the `no-boundary' criterion.} Since this analytic continuation gives rise to such features reminiscent of the original no-boundary wave function, it is worth exploring some features of the same. We shall now shortly see that, in addition to the aforementioned features, the state of primordial perturbations corresponding to the analytical continuation $a_0\rightarrow i/H$ is described by the Bunch-Davies state.
	
	In the limit $a_0\rightarrow i/H$, the on-shell background geometry is described by $\bar{q}_{+}(t)$, defined in \ref{onshell_nb_q}. Consequently, the roots of $\bar{q}_{+}$ are given by $\gamma_{nb}\equiv\gamma(i/H)=4i/(2i-p_1)$ and $\bar{\gamma}(i/h)=0$. On the other hand, the solutions $\chi_{\pm}(t)$ of the classical equations of motion of the perturbations reduces to:
	\begin{align}
	\chi_{\pm}(t)=\frac{1}{\sqrt{\bar{q}_{+}(t)}}\left[\frac{t}{t-\gamma_{nb}}\right]^{\pm\frac{(l+1)}{2}}\left[(1\mp(l+1))\gamma_{nb}+2(t-\gamma_{nb})\right]
	\end{align}
	The relevant solution, required by regularity near $t=0$, turns out to be $\chi_{+}(t)$. Note that this is equivalent to setting $\beta=0$. The on-shell action, corresponding to the Dirichlet conditions $\chi(1)=\chi_1$ and $\chi(0)=0$, can be evaluated to yield:
	\begin{align}
	S^{(2)}_{l}[\chi_{-},N_s(a_0=i/H)]&=\frac{V_3}{2}\frac{\bar{q}_{+}(1)}{N_s}\frac{\chi_{+}'(1)}{\chi_{+}(1)}\chi_1^2\\
	&=\frac{i l (l+2) \left(p_1^2+4\right) V_3 \chi_1^2}{4 H^2 (2 l-i p_1+2)}
	\end{align}
	Therefore, the total semi classical wave function $\tilde{\Psi}_{tot}(p_1,\chi_1)$, for the background and the perturbation, turns out to be:
	\begin{align}
	\tilde{\Psi}_{tot}(p_1,\chi_1)\propto e^{\frac{6V_3}{H^2\ell_p^2}} \exp\left[\frac{iV_3}{H^2\ell_p^2}\left( \frac{p_1^3}{24}+\frac{ p_1 }{2}\right)\right]\exp\left[-\frac{l (l+1) (l+2) \left(p_1^2+4\right) V_3 \chi _1^2}{2 \hbar H^2 \left(4 (l+1)^2+p_1^2\right)}-i\psi_l(p_1,\chi_1)\right]
	\end{align} 
	where the real phase $\psi_l(p_1,\chi_1)$ is given by
	\begin{align}
	\psi_l(p_1,\chi_1)=\frac{l (l+2) p_1 \left(p_1^2+4\right) V_3
		\chi _1^2}{4 H^2\hbar \left(4 (l+1)^2+p_1^2\right)}
	\end{align}
	Note that the perturbations are Gaussian distributed. In particular, in the $|p_1|\gg 1$ limit, the corresponding fluctuations are clearly described by a scale-invariant power spectrum.

	\section{Discussion}\label{sec_discuss}
	
	The large scale structures in our universe are believed to have originated from primordial quantum fluctuations. A promising realization of this idea is furnished by the inflationary scenario, wherein the primordial perturbations are magnified over time in an almost exponentially expanding universe. Usually, the relevant calculations are performed in the framework of ``quantum field theory in curved spacetime'', in which the spacetime describing the universe is treated classically, while the fluctuations on top of it are treated quantum mechanically. Here, we proposed a prescription-- in the minisuperspace model -- to perform the path integral over ``universes that are destined to expand''(UDE), which we promote here as a natural approach to quantum cosmology in the inflationary context. Surprisingly, the resulting wave function is found to be proportional to the Hartle-Hartle wave function. On the other hand, the saddle point geometry that dominates the path integral turns out to be pure de Sitter universe, hence establishing that our wave function may be viewed as the quantum extension of the de Sitter spacetime. 
	
	The quantum analysis of the background geometry also reveals that the wave function is invariant, apart from an overall phase fact, under the change of a certain positive real parameter $a_0$ that appears in the imposition of our initial condition. This, in turn, leads to the interpretation that our prescription is insensitive to the precise location, in time, of the initial space-like slice -- which is a desired feature. Motivated by this, we investigated whether $a_0$ can be analytically continued to a larger domain in complex $a_0-$plane, without changing the form of the wave function. We explicitly found this domain $\bar{\mathcal{D}}$, by demanding convergence of the path integral that defines our wave function and applying an extension of the Kontsevich-Segal theorem. Interestingly, two specific values of $a_0$, namely $\pm i/H$, having features reminiscent of the no-boundary proposal, lie on the boundary of $\bar{\mathcal{D}}$. It turns out that $a_0\rightarrow i/H$ exactly reproduces the Hartle-Hawking wavefunctions. This implies that by smoothly varying $a_0$ from $\mathbb{R}^{+}$ to $i/H$, one can analytically continue our prescription to one that realizes the no-boundary proposal. Incidentally, this smooth analytic continuation gives rise to a precise sense in which the no-boundary proposal is actually a complex extension of the sum over UDEs.
	
	Following that, we extend our analysis to include homogeneities, in the linear order in perturbation theory. The total wave function describing the background and the perturbation, that follows from our prescription, generates a scale-invariant power spectrum when the expansion rate is large enough. However, a straightforward implementation of our sum-over-UDEs approach, on its own, predicts only a one-parameter family of such states, rather than a unique one. We show that the corresponding parameter can be uniquely fixed by matching the wave function with that at the no-boundary point $a_0=i/H$. The state of perturbations thus obtained happens to be the Bunch-Davies vacuum. Thus, we have presented a concrete quantum cosmology extension of the conventional approach of quantum field theory in de Sitter space-time, that is often employed in the simple models of inflation. In particular, our study seems to confirm that the inflationary hypothesis, if implemented in a certain manner, predicts a family of scale-invariant power-spectrum for the perturbations. However, we can converge on a unique quantum state only if we include, as an extra ingredient, a version of the no-boundary proposal. 
	
	Before concluding, we would like to once again emphasize that the model presented here is proposed as a toy model to gain insight into the implications of an quantum cosmological treatment of inflation. If the inflationary scenario is a natural consequence of a certain more detailed quantum theory, one expects that, in an appropriately defined path integral, the corresponding measure may impart significant weight on universes that have an inflationary phase. The present work argues that, in the case of a theory with pure gravity and a positive cosmological constant, a simple initial condition can effectively capture the effect of microscopic details that prioritise the inflationary universes in a path integral sum. The most important prediction of this assumption is the scale-invariant power spectrum at late times. This is in sharp contrast to \cite{DiTucci:2019xcr}, where it was found that, in a different approach of studying inflation in quantum cosmology settings, the perturbations were found to be unsuppressed unless one assumes specially chosen initial conditions. However, the real consequences, i.e., those of interest to physical cosmology, of the current proposal will only emerge once it is extended to more realistic cases. For instance, it would be worth investigating how one can extend our approach from ``no-roll'', that we considered here, to more general inflationary scenarios. The starting point would be to appropriately extend the initial boundary condition considered in this work, which unambiguously filters universes with an expanding phase in the future, for a general model of inflation. We reserve this investigation for future work.  
	
	\section*{Acknowledgement}
	The author gratefully acknowledges illuminating remarks from Jean-Luc Lehners on an earlier draft of the paper. The author has also hugely benefited from several helpful discussions with Sumanta Chakraborty and  Vikramaditya Mondal.
	K.R. acknowledges support from the Research Associateship of the Indian Association for the Cultivation of Science (IACS), Kolkata, India. KR is currently supported by the Institute Postdoctoral Fellowship of Indian Institute of Technology Bombay, India. 
	
	\bibliography{Is_HH_wavefunction_the_quantum_dS}
	\bibliographystyle{utphys1}
\end{document}